\documentclass[fleqn,usenatbib]{mnras}

\usepackage{newtxtext,newtxmath}

\usepackage[T1]{fontenc}

\DeclareRobustCommand{\VAN}[3]{#2}
\let\VANthebibliography\thebibliography
\def\thebibliography{\DeclareRobustCommand{\VAN}[3]{##3}\VANthebibliography}

\usepackage{graphicx}	
\usepackage{amsmath}	

\usepackage{csquotes} 

\usepackage{caption}
\usepackage{subcaption} 

\usepackage[toc,page]{appendix} 
\usepackage{xcolor} 

\usepackage{hyperref} 

\graphicspath{ {./code_outputs/} }

\title[Particle acceleration in BH-XRB ejecta]{Cosmic rays, gamma rays and neutrinos from discrete black hole X-ray binary ejecta}

\author[N. Bacon et al.]{
Nicolas J. Bacon,$^{1,2}$\thanks{E-mail: njb207@cam.ac.uk}
Alex J. Cooper,$^{2}$
Dimitrios Kantzas,$^{3,4}$
James H. Matthews$^{2}$
and Rob Fender$^{2,5}$
\\
$^{1}$Institute of Astronomy, University of Cambridge, Madingley Road, Cambridge, CB3 0HA, UK\\
$^{2}$Astrophysics sub-Department, Department of Physics, University of Oxford, Keble Road, Oxford, OX1 3RH, UK\\
$^{3}$ Laboratoire d’Annecy-le-Vieux de Physique Théorique (LAPTh), USMB, CNRS, 74940 Annecy, France\\
$^4$Center for Astrophysics and Space Science (CASS), New York University Abu Dhabi, PO Box 129188, Abu Dhabi, UAE\\
$^{5}$Department of Astronomy, University of Cape Town, Private Bag X3, Rondebosch 7701, South Africa
}

\date{Accepted 2026 January 12. Received 2026 January 8; in original form 2025 October 8}

\pubyear{\the\year{}}

\begin{document}
\label{firstpage}
\pagerange{\pageref{firstpage}--\pageref{lastpage}}
\maketitle

\begin{abstract}
The origin of cosmic rays from outside the Solar system are unknown, as they are deflected by the interstellar magnetic field.
Supernova remnants are the main candidate for cosmic rays up to PeV energies but due to lack of evidence, they cannot be concluded as the sources of the most energetic Galactic CRs.
We investigate discrete ejecta produced in state transitions of black hole X-ray binary systems as a potential source of cosmic rays, motivated by recent $>100$ TeV $\gamma$-ray detections by LHAASO.
Starting from MAXI J1820+070, we examine the multi-wavelength observations and find that efficient particle acceleration may take place (i.e. into a robust power-law), up to $\sim2\times 10^{16}\mu^{-1/2}$ eV, where $\mu$ is the ratio of particle energy to magnetic energy.
From these calculations, we estimate the global contribution of ejecta to the entire Galactic spectrum to be $\sim1\%$, with the cosmic ray contribution rising to $\sim5\%$ at PeV energies, assuming roughly equal energy in non-thermal protons, non-thermal electrons and magnetic fields.
In addition, we calculate associated $\gamma$-ray and neutrino spectra of the MAXI J1820+070 ejecta to investigate new detection methods with CTAO, which provide strong constraints on initial ejecta size of order $10^7$ Schwarzschild radii ($10^{-5}$ pc) assuming a period of adiabatic expansion.
\end{abstract}




\section{Introduction}

Cosmic rays (CRs) are charged particles that propagate through space and bombard the Earth's atmosphere (e.g. \citealt{Beringer2012}).
CRs do not point back to their sources, as they are deflected by the magnetic fields in the intergalactic and interstellar media (ISM).
CRs are thought to be accelerated in blast-waves into a characteristic power-law spectrum in energy (e.g. \citealt{Longair2011}) by diffusive shock acceleration in magnetic fields (\citealt{Axford1977,Krymskii1977,Bell1978,Blandford1978,Drury1983,Begelman1984,Blandford1987}; for a review see e.g., \citealt{Matthews_2020}).
Galactic sources are thought to be limited to $\sim 10^{16}$ eV and below (e.g. \citealt{Matthews_2020}).
Supernova remnants (SNRs) are the main candidate for CRs below $\sim 10^{15}$ eV (originally proposed by \cite{Baade1934}, with strong evidence of pion decay indicative of CRs (likely protons) presented in \citealt{Ackermann2013}).
However, the lack of observations of TeV $\gamma$-rays (e.g. \citealt{Ahnen2017}) suggest that CR energies from SNRs do not reach the highest Galactic energies (PeV).
This is supported by theoretical work; e.g. \cite{Gabici2016} estimates only one SNR accelerating particles to $10^{15}$ eV at any time.
Hence, although there is likely a large flux of CRs from SNRs, the energies may not reach that of the knee in the CR spectrum at $10^{16}$ eV.

Black hole X-ray binaries (BH-XRBs, also referred to as microquasars) are BHs that accrete matter from a companion star and are observed in two distinct regimes: the \textquote{hard} state, which has a compact jet, and the \textquote{soft} state, which does not \citep{Remillard_2006}. 
The X-ray spectra for the two states are thought to be determined by the behaviour of the accretion disc and Comptonising corona (e.g. \citealt{Fender2004}).
In the transition from hard to soft, relativistic transient discrete ejections (ejecta, also referred to as \textquote{knots} or \textquote{blobs}) of plasma can be released ballistically \citep{Fender1997}, and ejecta may have energy on the high-end of estimates \citep{Fender2000,Carotenuto_2022}.
Ejecta are thought to be trans-relativistic blast waves which appear to be less energetic and less relativistic versions of $\gamma$-ray bursts \citep{Carotenuto2024,Cooper2025,Matthews2025}.
The release of ejecta is strongly associated with early-time radio flaring during hard-soft transitions, believed to be initially self-absorbed synchrotron emission from accelerated electrons \citep{Fender1997}.

XRB jets have been long-suggested as sites of diffusive shock acceleration and cosmic ray acceleration (e.g. \citealt{Romero2003,Romero2005, Bednarek_2005, Fender2005, Reynoso2008, Cooper_2020, Carulli2021, Kantzas2021, Kaci2025}), which is strongly supported by recent observations (e.g. \citealt{Aharonian2024,LHAASO2024}), however in this paper we consider discrete ejections from BH-XRBs, rather than hard-state jets.
Efforts to model particle acceleration in BH-XRB ejecta specifically are scarce and their contribution to the CR spectrum has been largely unexplored until recently (e.g. \citealt{Savard2025}).
CRs will interact with photons and thermal protons, producing pions which decay \citep{Kelner2006,Kelner_2008,Workman2022}.
The produced $\gamma$-rays and neutrinos point back to the source as these are not deflected by the Galactic magnetic field, unlike CRs.
If these emissions are detected, BH-XRB ejecta could be confirmed as a CR source.

In this paper, our aim is to determine the CR contribution from BH-XRBs through a combination of analytic calculation and analysis of recent observations, in particular for energies exceeding $10^{15}$eV.
In Section \ref{sec:acceleration} we lay out the fundamentals of CR acceleration and bulk ejecta dynamics and apply this to the recent MAXI J1820+070 ejection \citep{Bright_2020,Espinasse_2020,Wood2021}.
In Section \ref{sec:CRtotal}, we extend
these estimates to the population as a whole using radio flares to
estimate the total CR contribution.
Lastly, in section \ref{sec:NUG}, we explore $\gamma$-ray and neutrino emission from these ejections with both analytic and numerical calculations.

\section{Maximum cosmic ray energy}
\label{sec:acceleration}
\subsection{Maximum energy and ejecta expansion}
\label{sec:effc}
To calculate the contribution due to protons accelerated in ejecta from BH-XRBs to the CR spectrum, we first need to calculate the maximum proton energy attainable.
We assume that an ejection is composed of a magnetic field and equal numbers of protons and electrons, such that the ejection is charge neutral.
The composition of ejecta is unknown - there are observations of heavy hadrons (iron) in the jets of BH-XRB SS 433 \citep{Kotani1996,Migliari2002} and recent work \citep{Zdziarski2024} suggests that ejecta are electron-ion plasmas, but the possibility of electron-positron plasma has not been conclusively ruled out.
We focus on protons, however heavier ions could be accelerated to higher energies if they are present.
It has been shown \citep{Lagage1983,Hillas1984} that the characteristic maximum proton energy is
\begin{equation}
   \frac{E}{10^{15} \text{eV}} = \frac{D}{\text{pc}} \frac{B}{\text{$\mu$G}} \frac{\beta}{0.5}
\label{eq:simplemaxE}
\end{equation}
where the shock speed is $u=\beta c$, $B$ is the magnetic flux density and $D$ is the size of the accelerating region.

We assume the ejection is spherical with uniform energy density and uniform magnetic field strength for simplicity of calculations.
Additionally, we begin by assuming total energy (non-thermal particles and magnetic fields) is minimised, which is known as equipartition.
This assumption is relaxed later.
For time-varying luminosity $L_{\nu}(t)\propto\nu^{\alpha}$, $\alpha<0$ (e.g. \citealt{Longair2011}), the total internal energy $W_{\text{eq}}$ from equipartition (at which the ratio of $(\text{particle energy}/\text{magnetic energy}) \equiv\mu = 4/3$) satisfies:
\begin{equation}
W_{\text{eq}}(t) = \frac{7}{24\pi} V^{3/7} \left(6\pi G(\alpha) \eta L_{\nu}(t)\right)^{4/7}
\label{eq:equipartition}
\end{equation}
where $V$ is the volume and $G(\alpha)$ is a slowly-varying function of $\alpha$ (the spectral index of observed radiation) and the maximum frequency of emission.
$\eta$ determines the energy share between non-thermal protons and electrons.
Energy in the magnetic field $W_B \propto B^2 V$, which implies
\begin{equation}
   W_B \propto R^{9/7} L_\nu^{4/7} \implies B \propto R^{-6/7}L_\nu^{2/7}
\end{equation}
for a uniform magnetic field.
Hence the maximum energy scales as $\beta R^{1/7}L_\nu^{2/7}$ which should be similar for all ejections with a magnetic field so it is sufficient to analyse a snapshot of one event in detail here.

The total energy in non-thermal particles and magnetic fields as a function of time for MAXI J1820+070 is plotted in Fig. \ref{fig:Wt}.
We make the common assumption (e.g. \citealt{Persic2014}) that $\eta = 2$, which corresponds to equal energy in the non-thermal component of both species.
Since the minimum energies of the non-thermal spectra are likely different ($\approx$ rest-mass energy), there are different number densities of non-thermal protons and electrons. However, the jet may remain neutral, if charge neutrality is satisfied by the combined thermal (around $90\%$ of the total, see discussion below) and non-thermal components. 
The energy share between non-thermal electrons and protons is an open problem \citep{Marcowith2016,Matthews_2020} and can be varied, with one species having up to $\sim100$ times more energy than the other, e.g. \cite{Romero2008,Vila2011,Pepe2015}.
This would cause up to $\sim1$dex of variation in the total energy in non-thermal protons, roughly comparable to the variation due to the size uncertainty.
Some simulations (e.g. \citealt{Park2015,Crumley_2019}) suggest that protons are accelerated more efficiently, although these are not conclusive and assume unphysical proton-electron mass ratios.
The higher acceleration efficiency may be because electron gyroradii are smaller than proton gyroradii (for the same momentum) by a factor $\sqrt{m_\text{e}/m_\text{p}}$ (e.g. \citealt{Bell1978b}), although this is uncertain due to the difficulty in resolving gyroradii.
The issue is also discussed in \cite{Kantzas2023a} with respect to the hard state jets of BH-XRBs.
Furthermore, the system may not be in an equipartition state (i.e. $\mu$a not $4/3$); the effect on the proton energy spectrum is explored in Fig. \ref{fig:spectra}.

An alternative constraint can be placed on maximum proton energy $E_{\text{max}}$ by equating energy gains to synchrotron and adiabatic losses ($P\, dV$ work on the surroundings), which, when phrased in terms of reciprocal timescales, gives \citep{Cooper_2020}:
\begin{equation}
\frac{4}{3} \left(\frac{m_e}{m_p}\right)^2 c \sigma_T \frac{U_B}{m_p c^2} \frac{E_{\text{max}}}{m_p c^2} + \frac{\beta_{\text{exp}} c}{R} = \frac{\epsilon e c B}{E_{\text{max}}} c
\label{eq:full Et}
\end{equation}
where $U_B = W_B/V$ is the internal energy density in the magnetic field, $\beta_{\text{exp}} c$ is the lateral expansion speed, $\epsilon$ is the acceleration efficiency and the remaining symbols have their usual meanings.
Based on kinetic hybrid simulations (e.g. \citealt{Caprioli_2014}), $\epsilon \approx 0.1$ is appropriate for protons accelerated by non-relativistic shocks, applicable also in observations of non-thermal radiation \citep[e.g.][]{Morlino2012}.
For relativistic shocks, particle-in-cell simulations by \cite{Sironi2010} found $\epsilon \lesssim 0.3$ and similarly \cite{Crumley_2019} found $\epsilon \approx 0.1$ for shocks moving at $0.75c$.
We find that the proton synchrotron losses are negligible for our given parameters and hence
\begin{equation}
E_{\text{max}} = \frac{R \epsilon e B c}{\beta_{\text{exp}}} \sim 10^{16} \text{eV}
\label{eq:Et}
\end{equation}
This has the same dependence as the previously calculated maximum energy: $E_{\text{max}} \propto R B$.

Ejecta adiabatically expand over time from ejection ($t=0$), decreasing the magnetic field strength.
Ejecta are difficult to resolve due to the high angular resolution required, so it is not yet certain how long or how well they maintain their structure, although recent modelling work \citep{Cooper2025} finds a conical tophat jet without spreading appears to fit the data best.
Here, we model analytically the expansion as the ejection expands for constant magnetic flux $\phi = B R^2$, which physically equates to the magnetic field lines being frozen-in \citep{Longair2011}.
Magnetic energy follows 
\begin{equation}
W_B \propto B^2 V \implies W_B \propto {R^{-4}}{R^3} \propto R^{-1}
\label{eq:WBR}
\end{equation}
Individual particle energy satisfies \citep{Laan1966,Longair2011}
\begin{equation}
E \propto R^{-1} \implies
W_{\text{prot}} \propto R^{-1} \propto W_B
\label{eq:WPR}
\end{equation}
hence this is also an equipartition expansion.

\subsection{Results for MAXI J1820+070 ejection}
\label{sec:1820 results}

\begin{table}
	\centering
 \caption{Values of the model parameters.} \label{tbl2}
	\begin{tabular}{ccc} 
		\hline
Modelling parameter & Symbol & Value \\
  		\hline
Non-thermal $p$-$e^-$ energy share & $\eta$ & 2 (equal) \\
Non-thermal particle-B-field share & $\mu$ & 4/3 \\
Acceleration efficiency & $\epsilon$ & 0.1 \\
    \hline
Observational parameter & Symbol & Value \\
    \hline
Synchrotron spectral index & $\alpha$ & -0.6 \\
Ejecta size (AU) at 90 days & $R(90)$ & $6.2 \times 10^2 - 1.7 \times 10^4$ \\
Bulk ejecta Lorentz factor & $\Gamma$ & 1.7 \\
Viewing angle & $\theta$ & $64^\circ$ \\
		\hline
	\end{tabular}
\end{table}

\begin{table}
	\centering
 \caption{Calculated properties of the MAXI J1820+070 ejection 90 days \citep{Bright_2020} after launching for $\beta_{\text{exp}}=0.05$. The different sizes correspond to the lower and upper estimates.} \label{tbl1}
	\begin{tabular}{cccc} 
		\hline
$R(90)$ [AU]& $W_{\text{eq}}$ [erg]& Simple $E_{\text{max}}$ [eV] & Full $E_{\text{max}}$ [eV] \\
  		\hline
 $6.2 \times 10^2$ & $4.1 \times 10^{41}$ & $1.1 \times 10^{16}$ & $6.3 \times 10^{15}$ \\
 $3.3 \times 10^3$ & $3.4 \times 10^{42}$ & $1.4 \times 10^{16}$ & $8.0 \times 10^{15}$ \\ 
 $1.7 \times 10^4$ & $2.9 \times 10^{43}$ & $1.8 \times 10^{16}$ & $1.0 \times 10^{16}$ \\
		\hline
	\end{tabular}
\end{table}

In order to carry out an estimate for the total BH-XRB ejecta contribution to the CR spectrum, we begin by finding the maximum CR energy $E_{\text{max}}$ and internal energy $W_{\text{eq}}$ of the MAXI J1820+070 ejection.
Two apparently superluminal ejecta from MAXI J1820+070 were observed in 2017 \citep{Bright_2020}, with a re-analysis by \cite{Wood2021} finding a third slow-moving ejection.
These ejecta were observed by MeerKAT and eMERLIN simultaneously at the same frequency 90 days after initial radio flaring, allowing the size of the approaching ejection to be constrained to $6.2 \times 10^2\text{ AU}<R(90 \text{ days})<1.7 \times 10^4\text{ AU}$, using the difference in flux at two different angular resolutions (see also \citealt{Savard2025}).
Table \ref{tbl2} summarises the parameters used to derive the energetics presented in Table \ref{tbl1}.
The calculated $W_\text{eq}$ and size are in agreement with ejections from other BH-XRB sources \citep{Fender1999,Gallo2004,Brocksopp2007,Curran2014,Russell_2019,Cooper2025}.
We recalculate $W_{\text{eq}}$ utilising Chandra X-ray datapoints and assuming the ejection is optically thin throughout, as in \cite{Espinasse_2020} which found $\alpha = -0.6$ rather than $-0.7$.
We also account for relativistic beaming to find the true minimised total energy (e.g. \citealt{Savolainen_2010}; \citealt{Savard2025}), since energy $W_{\text{eq}}$ was originally calculated \citep{Bright_2020} for MAXI J1820+070 directly from the observed luminosity rather than the emitted luminosity.
For bulk Lorentz factor $\Gamma = \Gamma_{\text{min}} = 1.7$ as given by \cite{Bright_2020}, $W_{\text{eq}}$ increases by a factor 1.20 ($\alpha = -0.6$ and $\theta = 64^{\circ}$), although this may be an underestimate as the re-analysis by \cite{Wood2021} instead gave $\Gamma_{\text{min}} = 2.1$.
Observational uncertainties rule out using direct equations \citep{Miller-Jones2006} to solve for $\Gamma$ from the proper motion; recent work \citep{Saikia_2019,Carotenuto2024}; Cowie et al., in prep. suggests that a typical $\Gamma$ is 3, although $\Gamma$ may exceed this on kinematic grounds (\citealt{Fender2003,Fender2025,Lilje2025}).
Using $\Gamma = 3$ gives a boosted $W_{\text{eq}}$ 3.2 times greater for an inclination of $64^{\circ}$; correcting for beaming is especially important for more relativistic events.
We could also refine our estimate a little further: if we assume the shock is a forward shock, then we can relate $\Gamma$ to the Lorentz factor of the shock $\Gamma_{\text{sh}}$ \citep{Steiner_2012}.
For $\Gamma>1.7$, $\Gamma_{\text{sh}}>2.2$, bulk speed $\beta > 0.81$ and shock speed $\beta_{\text{sh}} > 0.89$.
The two velocities differ by $\lesssim 10\%$, so we take $\beta \approx \beta_{\text{sh}}$ to simplify our analysis.

\begin{figure}
	\includegraphics[width=\columnwidth]{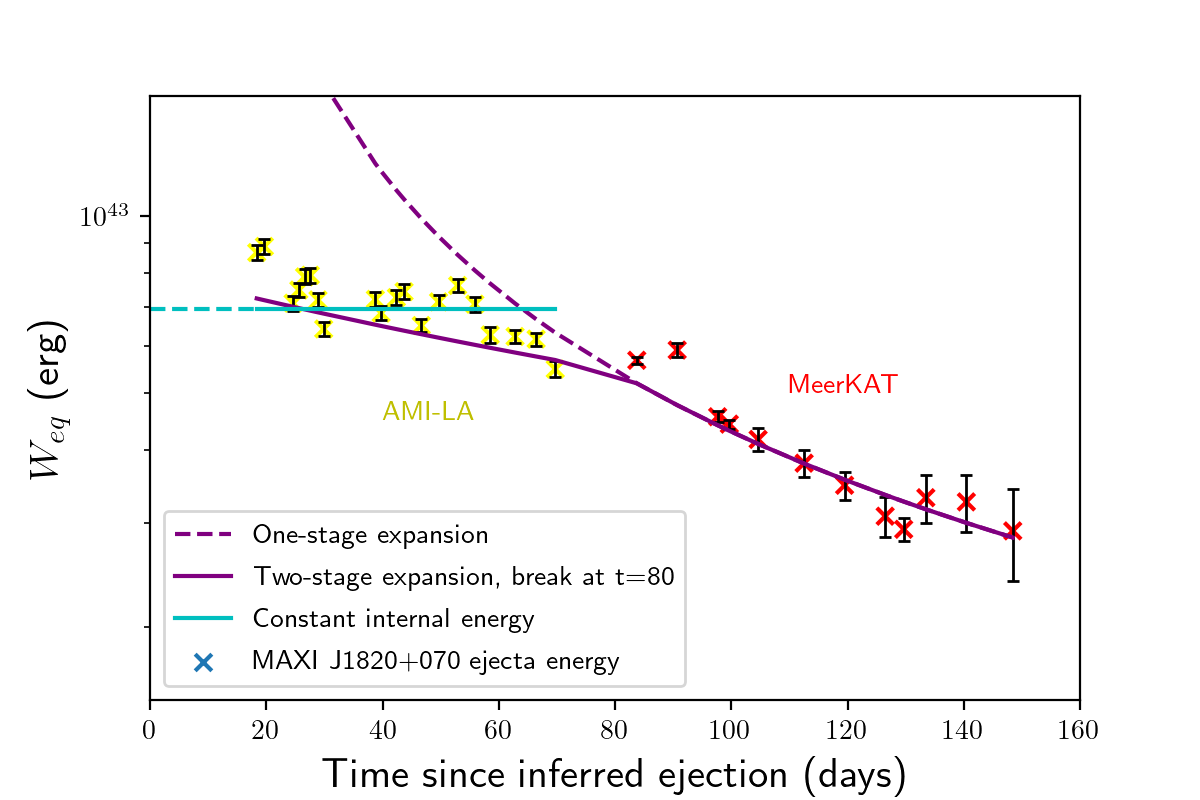}
\caption{Internal energy over time from equipartition for the MAXI J1820+070 ejection, as inferred from radio observations. An adiabatic expansion fits the data, with the purple dashed line highlighting the need for a break in expansion speed (best-fit at 80 days). Data plotted for $R(90 \text{ days})=3.3 \times 10^3$ AU.}
  \label{fig:Wt}
\end{figure}
\noindent

We model the lateral expansion of the ejection to be linear to estimate the expansion speed $\beta_{\text{exp}}c$ which affects the $E_{\text{max}}$ of protons attainable as in Eq. \eqref{eq:Et}.
For a full discussion, see Appendix \ref{app:integral}.
For an adiabatic expansion, the luminosity satisfies $L_{\nu} \propto R^{-2p}$ (e.g. \citealt{Longair2011}).
A Taylor–von Neumann–Sedov expansion \citep{Sedov1946} is appropriate for the late-time behaviour ($t\gtrsim80$d) when the bulk dynamics become non-relativistic.
The internal energy varies slowly at these times, so we label this stage as a pseudo-steady state.
After fitting $L_{\nu} \propto R^{-2p}$ to the data with $R(t) = R_0 + \beta_{\text{exp}} c t$ (using the geometric mean size $R(90 \text{ days})=3.3\times10^3$ AU, with the only free parameter being $\beta_{\text{exp}}$), we find that there are two regimes for $\beta_{\text{exp}}$ with a best-fit break at $t \approx 80$d, consistent with \cite{Bright_2020}\footnote{Note that this may correspond to the reverse shock crossing timescale (e.g., \citealt{Cooper2025,Matthews2025,Savard2025})}.
We then use Eq. \eqref{eq:equipartition} to calculate the internal energy $W_{\text{eq}}$ over time, plotted in Fig. \ref{fig:Wt}.

\begin{figure}
	\includegraphics[width=\columnwidth]{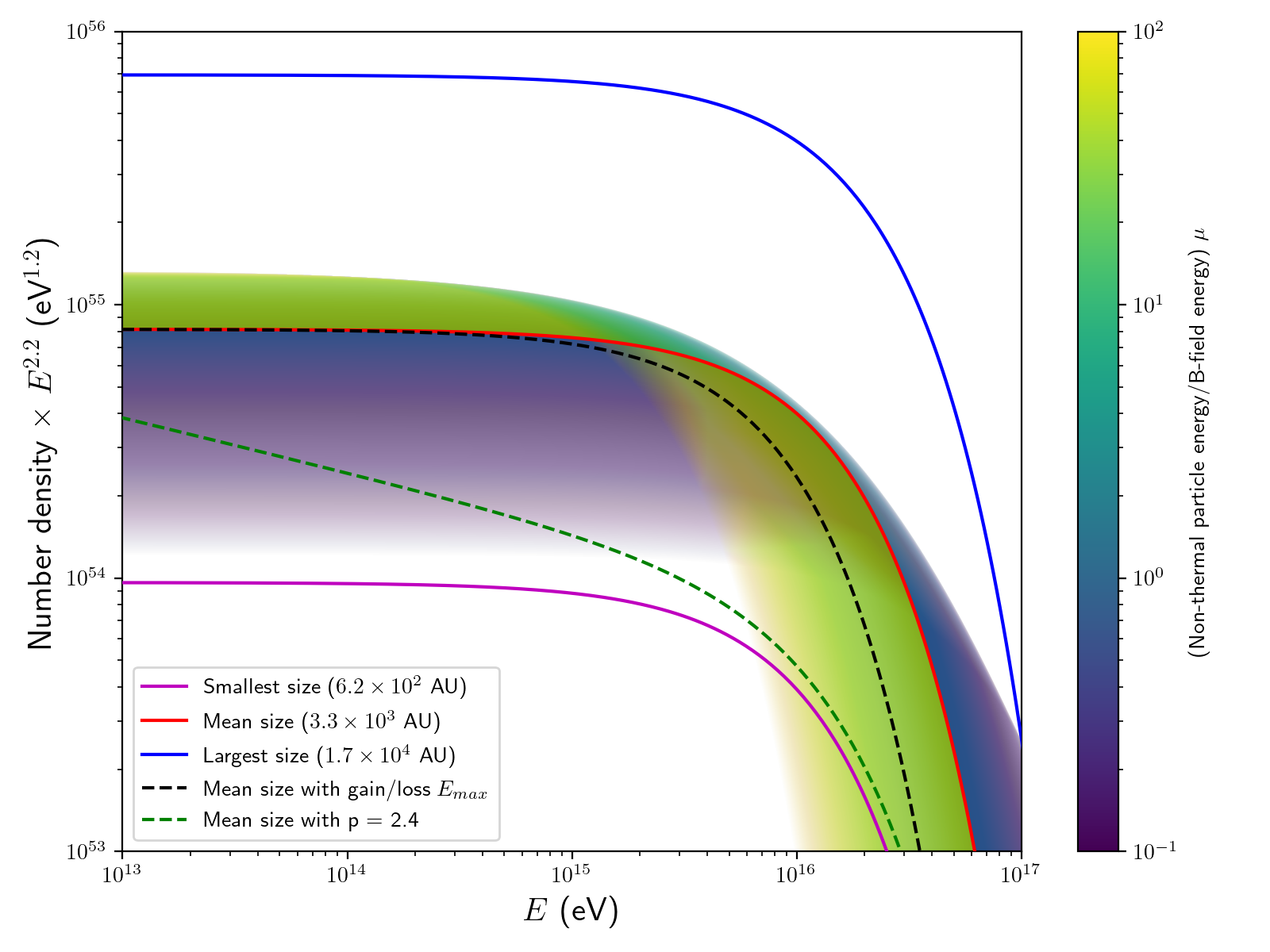}
\caption{The proton energy spectrum of the MAXI J1820+070 ejection at $t=90$ days after inferred ejection, multiplied by volume. The shaded area represents departures from equipartition ($0.1<\mu<100$) for the mean size. Maximum energy scales as $E_\text{max}\propto \mu^{-1/2}$ and in general size uncertainty of MAXI J1820+070 at 90 days dominates over potential deviations from equipartition.}
  \label{fig:spectra}
\end{figure}

We assume an energy spectrum of protons accelerated by shocks of 
\begin{equation}
N(E) dE = N_0 E^{-p} \exp \left({-\frac{E}{E_{\text{max}}}}\right) dE 
\end{equation}
with power-law index $p = -2\alpha + 1$ (spectral index $\alpha < 0$; e.g. \citealt{Matthews_2020}).
For the total energy in non-thermal protons we can write:
\begin{equation}
W_{\text{prot}} = \int_{E_{\text{min}}}^{\infty} E N(E) \,dE = \frac{4}{7}W_{\text{eq}} \times \frac{\eta-1}{\eta}, \text{ } \eta > 1
\label{eq:Wprot}
\end{equation}
These calculations (see Appendix \ref{app:integral}) give a lower bound $E_{\text{min}}$ of 1 GeV for $p=2.2$, in agreement with conservative estimates $\equiv m_\text{p} c^2$ from \cite{Kantzas2023a}.
Other similar works (e.g. \citealt{Romero2003,Romero2008,Vila2011,Pepe2015}) suggest that the minimum energy of non-thermal protons should be between $2-120$ GeV.
Note that the overall number of non-thermal protons is relatively insensitive to minimum energy: spectrum normalisation $N_0 \propto E_\text{min}^{p-2}$.
If we take $E_\text{min}=10$ GeV, the normalisation rises by a factor 1.6; for $E_\text{min} = 100$ GeV it rises by a factor 2.5.

The approaching ejection has spectral index $\alpha = -0.59 \pm 0.01$ and the receding ejection has $\alpha = -0.65 \pm 0.01$ (using Chandra data - see \citealt{Espinasse_2020}) so $p_{\text{app}}=2.2\pm 0.02$ and $p_{\text{rec}}=2.3\pm 0.02$.
We take the power-law index $p=2.2$ for the spectra; in general, $2<p\lesssim 2.6$ and shock theory tends to predict $p = 2.2-2.3$ (e.g. \citealt{Drury1983}).
The number density of accelerated protons as a function of energy for MAXI J1820+070 ejection is shown in Fig. \ref{fig:spectra}.
The larger the size, the greater the number of accelerated protons across the spectrum.
The size is the most important factor in determining the CR power, even when accounting for departures from equipartition.
\cite{Schekochihin2009} suggests that a reasonable range for the energy partition between non-thermal particles and magnetic fields is $0.1 < \mu <100$; below $\mu \sim 0.1$ maximum energy estimates become quite large due to the $\mu^{-1/2}$ scaling.
Departures from equipartition will change the shape of the CR energy spectrum, as in Fig. \ref{fig:spectra}, but even extreme values of $\mu$ produce variations less than both size uncertainty and differences between events, as discussed in Section \ref{sec:CRtotal}.
This highlights the importance of multi-frequency observations of ejecta, and the need for higher spatial resolution where possible.
We set the $E_{\text{max}}$ to be the simpler estimate as in Eq. \eqref{eq:simplemaxE}; there is a small difference for $E_{\text{max}}$ determined by adiabatic losses at energies $\gtrsim 10^{15}$ eV.
For $p=2.4$ \citep{Bright_2020} there are more low-energy and fewer PeV protons but again the size uncertainty is dominant.
Regardless of size, a large number of protons are accelerated to energies greater than $10^{15}$ eV, with a robust CR $E_{\text{max}} = 1.6\times10^{16} \mu^{-1/2}$ eV as smaller sizes fit the data better (see Appendix \ref{app:expansion} for details).

\section{Contribution to the cosmic ray spectrum}
\label{sec:CRtotal}
We now estimate the global contribution to the CR spectrum from discrete ejections.
If there is limited variability in energetics and behaviour between different ejections, BH-XRBs could partially explain the findings of \cite{Cao2024} which suggests the knee in the CR spectrum is due to protons.
Additionally, this would provide further theoretical backing to recent persistent high energy ($\sim$100 TeV) $\gamma$-ray observations in the vicinity of MAXI J1820+070 and four other recently active or persistent BH-XRBs \citep{LHAASO2024}.
Repeating the calculations above for the ejection from XTE J1908+094 \citep{Rushton2017} gives  $E_{\text{max}} \approx 1\times10^{16} \bmu^{-1/2}$ eV, which suggests the MAXI J1820+070 ejection is not an abnormal event.

Energetics estimates for most other ejecta cannot be performed in the same manner as MAXI J1820+070 due to the lack of strong size constraints.
Instead, we proceed utilizing the better characterised energy of self-absorbed radio flares.
The energy in flares $W_{\text{flare}}$ is calculated by Cowie et al., in prep., using the framework in \cite{Fender_2019}) which assumes the observed radio flares are associated with a transition from optically thick to optically thin, and that the bulk Lorentz factor of each region $\Gamma$ = 3.
Flares have been strongly associated with ejection (e.g. \citealt{Fender2009}), with a flare coincident with all documented ejections, and are much more easily observed.
These flares may correspond to the same material as the large-scale ejecta, with absorbed radiation stemming from particles accelerated prior or during the ejection event.
It is possible to estimate the ejecta size as $\beta_{\text{exp}} c t_{\text{flare}}$, where $t_{\text{flare}}$ is the rise time of the flare (e.g. \citealt{Fender2023}).

\begin{figure}
	\includegraphics[width=\columnwidth]{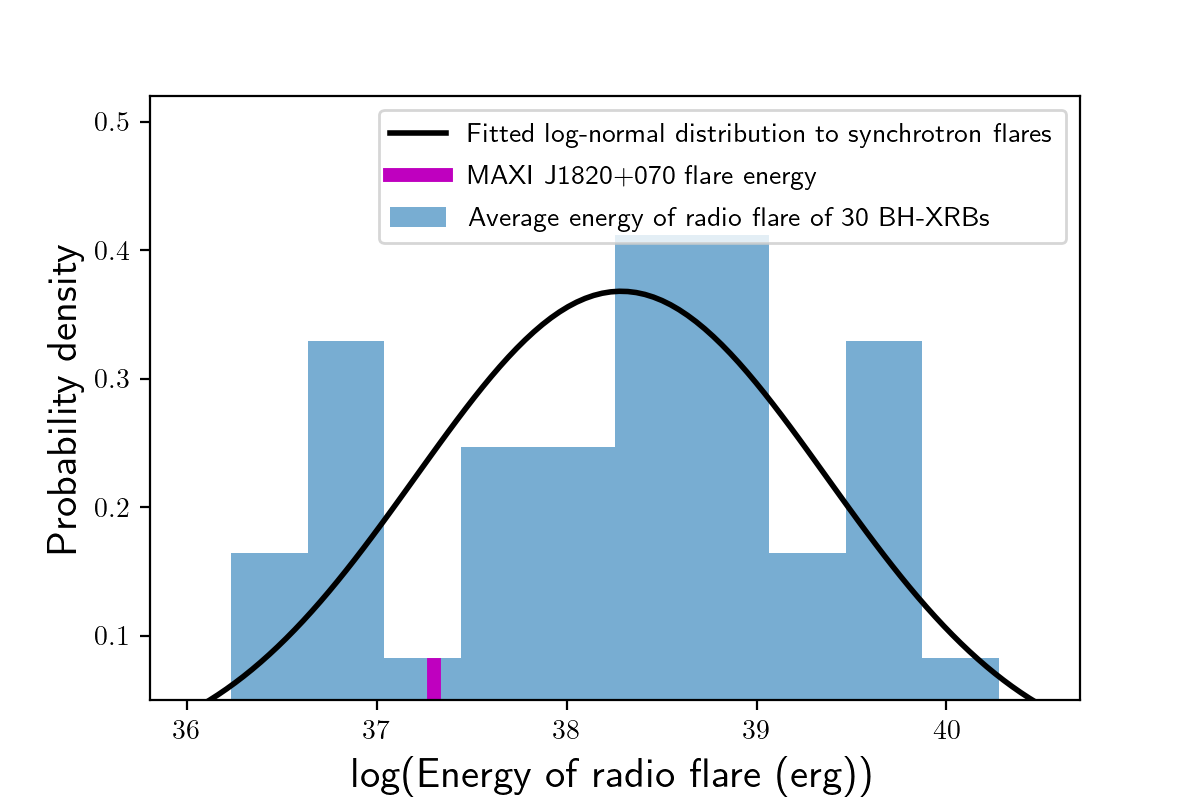}
\caption{Fitted log-normal distribution of synchrotron flare energies of 30 candidate BH-XRBs (Cowie et al., in prep.), with the MAXI J1820+070 flare energy indicated. The mean flare energy is $2\times10^{38\pm1.1}$ erg. Assuming a proportional relation between flare and ejecta energies gives an internal energy distribution for ejecta.}
  \label{fig:flaring}
\end{figure}

Ejecta are not always detected after a flare (e.g. \citealt{Fender2009}); possibly because flares are isotropic whereas ejecta have undetectable fluxes due to beaming.
Alternatively, these radio flares may be associated with a different phenomenon, e.g. slower-moving ejecta \citep{Wood2021} or other changes in accretion.
Nevertheless, we assume that each flare corresponds to a canonical fast-moving ejection with internal energy $W_{\text{ej}}$, and additionally that the energies are directly proportional:
\begin{equation}
\frac{W_{\text{ej}}}{W_{\text{flare}}} \equiv \kappa
\label{eq:Wprop}
\end{equation}
where $\kappa$ is a constant for all events; an oversimplification due to lack of data, but appropriate as we expect more energetic ejecta to be associated with stronger flaring events.
Note that $W_{\text{ej}}$ is time-dependent as in Fig. \ref{fig:Wt}, particularly close to ejection, so we take $W_{\text{ej}}$ at observation.
Other relationships are possible, for instance we could apply Eq. \eqref{eq:equipartition} directly, however Eq. \eqref{eq:Wprop} is sufficient to roughly estimate the BH-XRB population contribution to the CR spectrum.
With Eq. \eqref{eq:Wprop}, we relate the distribution of flare energies in Fig. \ref{fig:flaring} (to which we fit a log-normal distribution) to a distribution of internal ejecta energies $W_{\text{ej}}$.

As well as MAXI J1820+070, we examine ejections from XTE J1908+094 \citep{Rushton2017}, MAXI J1535-571 \citep{Chauhan2019,Russell_2019,Cooper2025} and prolific GRS 1915+105 \cite{Tetarenko_2016}.
There are two separate estimates for the MAXI J1820+070 flare energy in \cite{Bright_2020}; we choose to use the method presented in \cite{Fender_2019} to calculate the flare energy, which is the method used for the other ejections in Cowie et al., in prep.
Another relevant event is the ejection from recently-discovered MAXI J1348–630, however for this ejection only the bulk kinetic energy of the ejecta $(\Gamma - 1) M c^2$ is estimated \citep{Carotenuto_2022} rather than $W_{\text{ej}}$.
We estimate the internal energy of MAXI J1348–630 by assuming the ratio of kinetic energy to $W_{\text{ej}}$ is the same as for MAXI J1820+070.
This gives $(\Gamma - 1) M c^2 \approx 10 W_{\text{ej}}$, in agreement with \cite{Fender1999,Gallo2004} and the discussion of $\epsilon$ in Section \ref{sec:effc}.
We can also consider the Eddington luminosity, $L_{\text{Edd}} = 1.3\times10^{38} M_{\text{BH}}/M_\odot$ erg/s (see e.g. \citealt{Longair2011}).
Luminosity can exceed $L_{\text{Edd}}$ in non-steady state behaviour (e.g. \citealt{Done2004}), but flaring tends to occur at and above $\approx 0.1L_{\text{Edd}}$.
According to the WATCHDOG catalogue \citep{Tetarenko_2016}, the average luminosity of GRS 1915+105 is $6.3\times10^{38}$ erg/s $= 0.4L_{\text{Edd}}$.
Hence
\begin{equation}
    \text{Summed GRS 1915+010 ejecta energies} \lesssim \int 0.4L_\text{Edd} \,dt
\end{equation}
such that the average power is below 0.4 times the Eddington luminosity.
GRS 1915+010 had 486 flares over a 4000 day observation period with a mean flare energy of $1.4\times10^{40}$ erg (Cowie et al., in prep.), giving $\kappa < 5\times10^4$.
This inequality could be violated if ejections are significantly more energetic than steady-state behaviour.
The different values of $\kappa$ are summarised in Table \ref{tbl_kappa}.
Constraining all $\kappa < 5\times10^4$ and taking a weighted mean average gives $\log\kappa = 4.1\pm0.4$.

\begin{table*}
	\centering
 \caption{Values of $\kappa$, the ratio of internal over flare energy, as in Eq. \eqref{eq:Wprop}, for different events. Uncertainties in ejecta size and flare energy (Cowie et al., in prep.) dominate.} \label{tbl_kappa}
	\begin{tabular}{cccc} 
		\hline
		BH-XRB & Calculated $\kappa$ & $\log\kappa$ & Notes \\
		\hline
MAXI J1820+070 & $1\times10^4\lesssim \kappa \lesssim 8\times10^5$ & $4.0\lesssim \log\kappa \lesssim 5.9$ & Flare is $e^-$ synchrotron emission \citep{Bright_2020,Espinasse_2020} \\
XTE J1908+094 & $4\times10^2\lesssim\kappa \lesssim3\times10^4$ & $2.6\lesssim \log\kappa \lesssim 4.5$ & $W_{\text{ej}}$ smaller than most other events \\
MAXI J1535-571 & $1\times10^4\lesssim\kappa \lesssim1\times10^5$ & $4.1\lesssim \log\kappa \lesssim 5.0$ & Size uncertainty due to ejecta unresolved mitigated by modelling \\
MAXI J1348–630 & $8\times10^2\lesssim \kappa \lesssim 5\times10^4$ & $2.9\lesssim \log\kappa \lesssim 4.7$ & Take internal energy fraction $= 0.1$ and cavity density $= 0.0010 \text{ cm}^{-3} $ \\
GRS 1915+10 & $\kappa < 5\times10^4$ & $\log\kappa < 4.6$ & Ejecta energy is Eddington-limited: luminosity cannot exceed $0.4L_{\text{Edd}}$ \\
		\hline
	\end{tabular}
\end{table*}

\begin{figure}
	\includegraphics[width=\columnwidth]{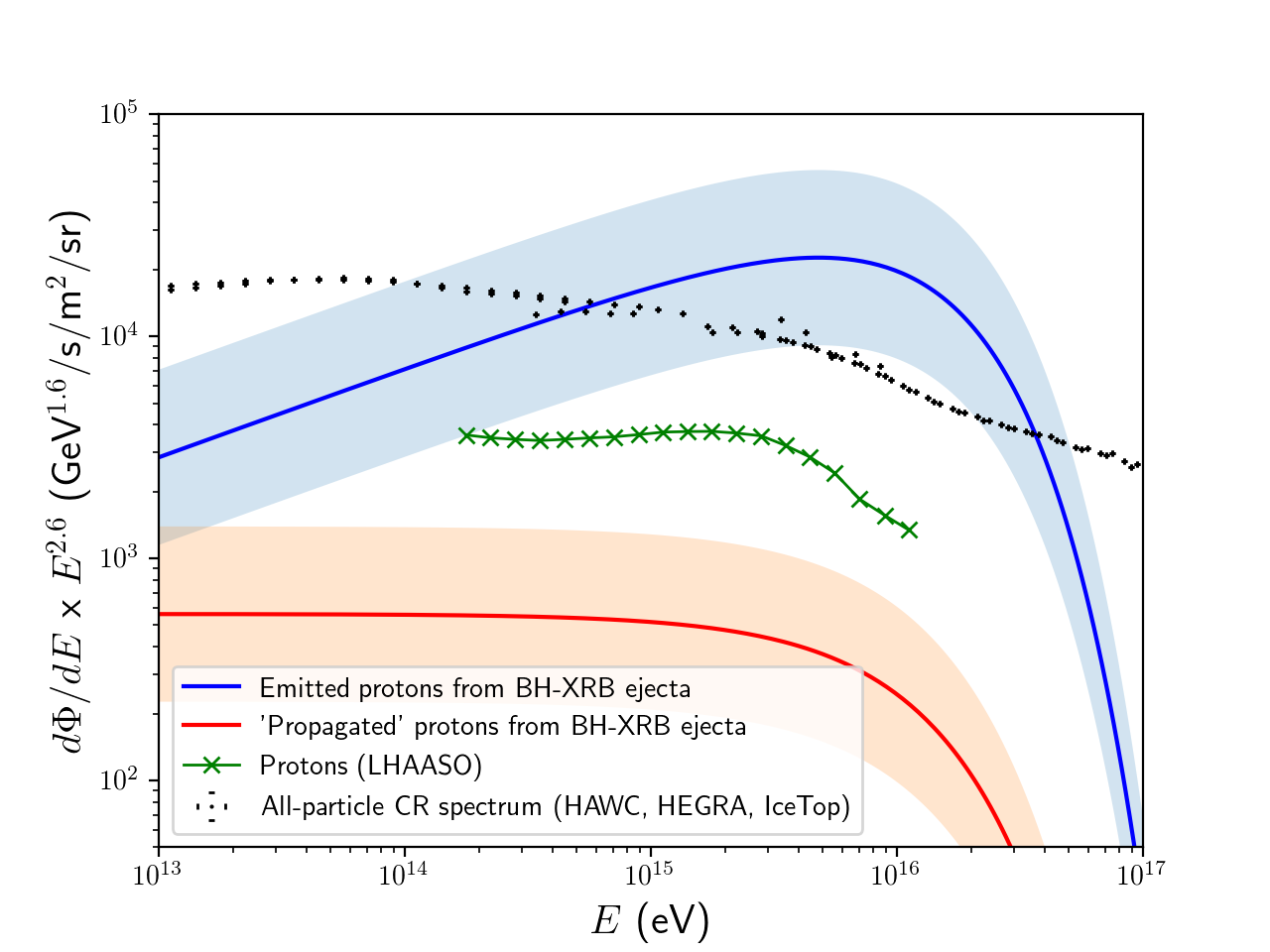}
\caption{Blue: predicted contribution to the CR spectrum with $p=2.2$, $E_{\text{max}} = 10^{16}$ eV and $\kappa = 1.3\times10^4$; red: a rough estimate (see discussion below) for the effect of propagation, with $p=2.6$, $E_{\text{max}} = 10^{16}$ eV and $\kappa = 1.3\times10^4$, normalised to have the same energy in CRs; data: CR spectrum \citep{Maurin2023} and protons \citep{LHAASO2025}.}
\label{fig:CRs}
\end{figure}

Sampling from the log-normal distribution of flares in Fig. \ref{fig:flaring} gives a mean flare energy of $\approx 4\times10^{39}$erg.
According to BlackCAT data (\citealt{CorralSantana2016}), 73 candidate stellar-mass BH-XRBs have been observed in the Milky Way.
Analysis of the spatial distribution implies there are $\sim1300$ total transient BH-XRB systems in the Galaxy \citep{CorralSantana2016}.
We estimate that the combined BH-XRB flaring activity is equivalent to 10 active BH-XRBs in the Milky Way that flare as frequently as GRS 1915+105, roughly one flare/day (Cowie et al., in prep.), analogous to \cite{LHAASO2024} who estimate the proton luminosity from Galactic BH-XRBs as $\gtrsim$ 10 times that of SS 433.
Imposing the upper limit on $\kappa$ from GRS 1915+10, we find the CR luminosity $L_{\text{CR}}$ satisfies $3.0\times 10^{38} \lesssim L_{\text{CR}} \lesssim 1.9\times 10^{39}$ erg/s, which is 0.2-1.2\% of the total Galactic luminosity of $1.5\times 10^{41}$ erg/s \citep{Dar_2001}, in approximate agreement with the hard-state contribution found by \cite{Cooper_2020}.
Although ejecta are not significant for the entire CR spectrum, their contribution may become important at the knee (Fig. \ref{fig:CRs}).
The contribution would be slightly higher if we take $W_\text{ej}$ at early times instead: for MAXI J1820+070, the energy halves between early times and observation according to Fig. \ref{fig:Wt}, increasing the maximum contribution to $\sim$ 2\%.
Additionally, as in Fig. \ref{fig:CRs}, this contribution peaks at PeV energies and suggests that BH-XRB ejecta could alone account for the knee, in agreement with \cite{LHAASO2024} from modelling of SS 433 and the analysis in \cite{Kaci2025}.

There are two important considerations to make on this estimation of overall CR power.
Firstly, CRs travel for a distance $d$ and time $t$ related by \citep{Aharonian2024}:
\begin{equation}
d \approx \sqrt{4 D_{\text{gal}} t}
\end{equation}
where $D_{\text{gal}} = 10^{28} \left(\frac{E}{1 \text{GeV}}\right)^{\frac{1}{3}}$ cm$^2$ s$^{-1}$ is the Galactic diffusion coefficient.
This gives a CR travel time of $\approx 10^{6}$ years from Galactic sources.
Since the population of BH-XRBs should be roughly correlated to the star formation rate (SFR) (e.g. \cite{Grimm2003} for high-mass XRBs), we assume that the BH-XRB population should be fairly constant, given that the SFR in the Milky Way has barely changed in the last $4\times 10^{9}$ years (e.g. \citealt{Revaz2016}), suggesting a comparison between observed BH-XRB activity and current CR data is valid, notwithstanding the possibility of a single, nearby dormant source which may skew current data.
Secondly, we haven't considered the effect of propagation - without this, BH-XRB ejecta could entirely account for the knee.
This phenomenon has been studied and modelled in depth by \cite{Evoli_2017}, and the general effect is to steepen the power-law (so that the $p\approx2.2$ predicted by shocks is changed to $p=2.6-2.7$ as in the CR spectrum) and $E_{\text{max}}$ decreases.
The full calculations and Monte Carlo simulations are beyond the scope of this paper, see however \cite{Cooper_2020} and \cite{Kantzas2023c} for calculations involving CRs from the hard state; we investigate the propagation of CRs from ejecta in a future work.
In Fig. \ref{fig:CRs} a curve with $p=2.6$ and equal energy normalisation is plotted to roughly emulate the result.
The diminished knee contribution instead suggests that BH-XRB ejecta only partially account for the PeV Galactic CRs ($\sim$ 5\%).
Increasing $\mu$ to $100$ increases the upper limit to $\sim 15\%$ of Galactic CRs from BH-XRB ejecta.
Further observational and theoretical work is needed to quantify the exact contribution.

\section{Emission of gamma-rays and neutrinos}
\label{sec:NUG}
As well as producing PeV CRs, ejecta will also emit $\gamma$-rays and neutrinos from non-thermal protons interacting with the rest of the ejecta.
Recently, a number of BH-XRBs have been detected with extended $>10-100$ TeV $\gamma$-ray emission (e.g. \citealt{Romero2005,Kantzas2021,LHAASO2024}) and possibly neutrinos too \citep{Cooper_2020,Kantzas2023c}, suggesting they are accelerating particles to $\sim$PeV energies.
The connection between the multi-wavelength observations of discrete ejecta and the TeV gamma-ray detections is not immediately clear.
On the one hand, we and others \citep{Savard2025} have shown that the discrete ejecta can plausibly reach up to and beyond PeV energies, and ejecta emission has been linked \citep{LHAASO2024}.
On the other hand, the $\gamma$-ray detections are not localised to the BH-XRBs; observations by \cite{LHAASO2024} subtend angles of $\sim0.5^\circ$.
In this section, we calculate fluxes of $\gamma$-rays and neutrinos from the MAXI J1820+070 ejection and examine if these are detectable and hence responsible for the TeV $\gamma$-rays observed.

When high energy protons collide with photons or with other protons, there is a probability to produce pions ($\pi$) (see e.g. \citealt{Workman2022}).
Production of $\pi$ is maximum when the particle density and radiation field strength are high but the ejection has become large enough to have a high maximum proton energy.
Our method to calculate an output spectrum of $\gamma$-rays or neutrinos from proton-proton interactions ($p+p$) or proton-photon interactions ($p+\gamma$) is as follows.
Taking the example $p+p\rightarrow p+p+\pi^{0}, \text{ }\pi^{0}\rightarrow \gamma \gamma$, we start by making the delta-function approximation that energy of pion $E_{\pi} = A_{\pi} E_{\text{kin}}$ where $E_{\text{kin}}$ is the kinetic energy of the proton and $A_{\pi} \approx 0.17$ is the \textquote{inelasticity} and is a constant (e.g. \citealt{Reynoso_2008}).
Making the relativistic approximation $E_{\text{kin}} = E_{p}$, the output spectrum of pions is
\begin{equation}
q_{\pi}(E_{\pi}) \approx cn_{H} \int \delta(E_{\pi} - A_{\pi} E_{p}) \sigma_{pp}(E_{p}) N_{p}(E_{p}) \,dE_{p}
\end{equation}
where $q_{\pi}(E_{\pi})$ is the number of pions produced per unit time, volume and energy, $\sigma(E)$ is the cross-section and $N_{p} (E_{p})$ is the number density of protons, both as a function of energy.
The spectrum is proportional to the number density of thermal protons $n_H$.
Integrating:
\begin{equation}
q_{\pi}(E_{\pi}) \approx cn_{H} \sigma_{pp}\left(\frac{E_{\pi}}{A_{\pi}}\right) N_{p}\left(\frac{E_{\pi}}{A_{\pi}}\right)
\end{equation}
The spectrum of pions $q_{\pi}(E_{\pi})$ can then be used to calculate the output $\gamma$-rays or neutrinos for various parametrisations of the cross-section $\sigma(E)$.
In the case of $\gamma$-rays, the output spectrum $q_{\gamma}(E_{\gamma})$ is
\begin{equation}
q_{\gamma}(E_{\gamma}) \approx 2 \int_{E_\text{thresh}}^{\infty} \frac{q_{\pi}(E_{\pi})}{E_{\pi}} \,dE_{\pi}
\end{equation}
This gives a $\gamma$-ray spectrum with a spectral index similar to the non-thermal protons and decreased by a factor $\approx A_{\pi}$.
Full derivations can be found in \cite{Kelner_2008} ($p+\gamma$) and \cite{Kelner2006} ($p+p$).
The above arguments provide us with the theoretical foundations to find expected $\gamma$-ray and neutrino emissions.

For $p+p$ interactions, we take the target proton field to be the thermal protons of the ejecta, of which there are roughly ten times as many as non-thermal protons (Section \ref{sec:effc}).
For $p+\gamma$ interactions, the target photon field is taken to be the synchrotron spectrum from \cite{Espinasse_2020}.
We assume $E_{\text{max}}$ of photons $\sim$ $E_{\text{max}}$ of electrons.
Electron $E_{\text{max}}$ is calculated via two methods: firstly, with the gain/loss estimate Eq. \eqref{eq:full Et} ($E_{\text{max}} = 8\times 10^{10}$ eV), and secondly neglecting adiabatic losses and setting $t_{\text{sync}}$ = 90 days ($E_{\text{max}} = 1\times 10^{12}$ eV).
Although both of these are likely to be overestimates, for $E_{\text{max}}\gtrsim 10^{8}$ eV, there is no discernible difference to the output spectrum.
We assume that the synchrotron spectrum a year after ejection is the same as at $t = 90$ days, and that the conditions of the ejecta don't change significantly over timescales on the order of years, valid assuming conservation of magnetic flux $\phi$.
The remaining required parameters - mass of BH ($8.5M_\odot$), distance to BH-XRB (2.96 kpc) and inclination angle ($64^\circ$) - are taken from \cite{Atri2020} and \cite{Torres2020}.
We will now use these observations and assumptions in combination with the above theoretical groundwork to explicitly calculate expected $\gamma$-ray and neutrino spectra.

Previously, it was unknown whether $\gamma$-rays or neutrinos could be detected from ejecta, however according to the simulations, the rate of production in steady state is far too low, unlike the hard state jets which could produce measurable $\gamma$-ray flux \citep{Kantzas2021}.
The spectra are presented in the Appendix \ref{app:steady state} in Fig. \ref{fig:gamma-rays} and Fig. \ref{fig:neutrinos} and show that $\gamma$-rays and neutrinos at $t = 90$d are unobservable, in agreement with $\gamma$-ray observations by \cite{Hoang2019,Abe2022}.
Furthermore, the spectra suggest that $p+p$ interactions are dominant over $p+\gamma$, assuming the $\gamma$ come from electron synchrotron emission only and have the observed power law flux.
Even with upper limits by taking ejecta radius and distance to the BH-XRB to be the most favourable within $1\sigma$ uncertainties, as well as allowing for departures from equipartition (away from $\mu$ = $4/3$, as in \citealt{Schekochihin2009}), the emitted $\gamma$-ray and neutrino flux would be orders of magnitude too small to be detected even over long observation times.
We can do the same calculations for earlier times, using Fig. \ref{fig:Wt} for the internal energy and radius over time.
The closer to ejection, the higher the rate of $\gamma$-ray and neutrino production due to the increasing energy density.
However, the energy density is still far too low to produce detectable rates at astrophysical distances.
These new calculations suggests that $\gamma$-ray and neutrino observation in the pseudo-steady state is highly unlikely.
It is possible that repeated ejections, perhaps combined with particle diffusion in the ISM may explain the TeV $\gamma$-ray emission \citep{LHAASO2024}, but this requires the particles to propagate a large distance without cooling significantly.
Further work on the propagation of CRs and bulk ejecta dynamics is needed to definitively link ejecta to these emissions.

\begin{figure}
    \centering
     \includegraphics[width=\columnwidth]{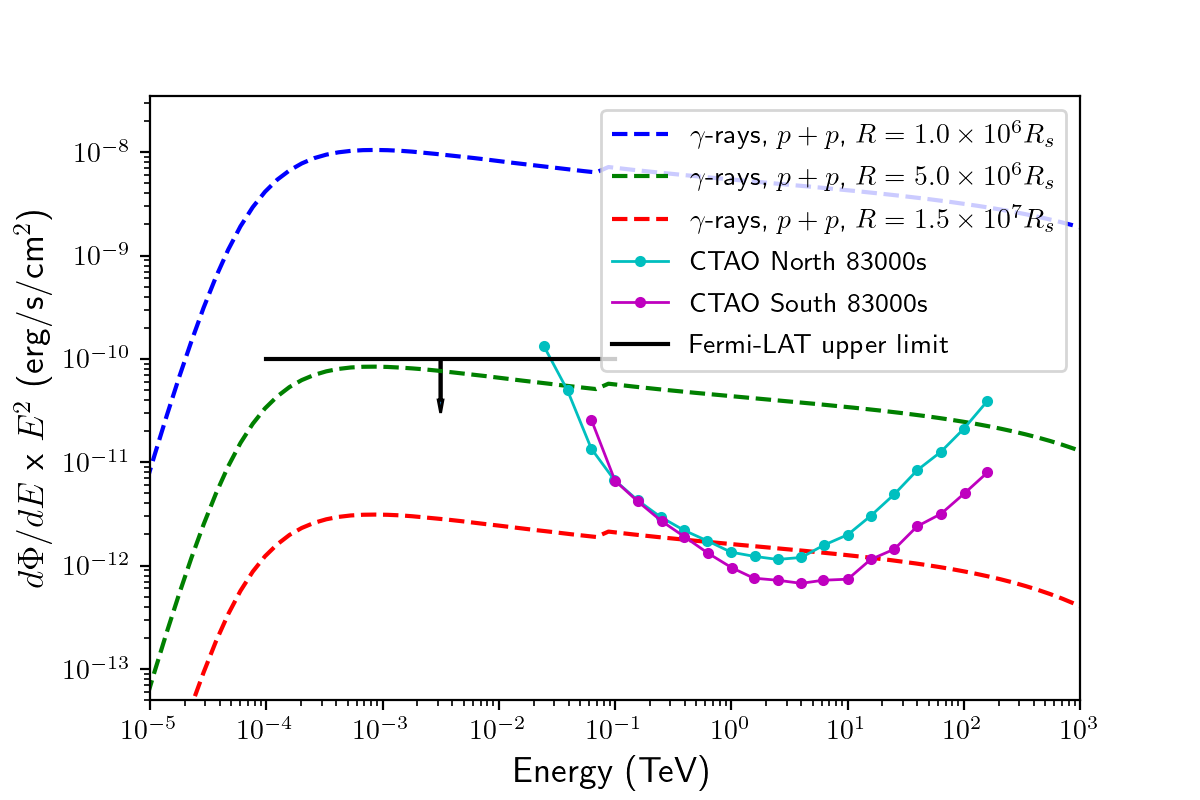}
    \caption{$\gamma$-rays (from $p+p$ interactions in MAXI J1820+070) could have been detectable with CTAO \citep{Observatory2021} for ejecta radii $\lesssim 1.5\times10^7 R_S$.} 
\label{fig:early gamma-rays}
\end{figure}

It is interesting to see what is predicted by our model if we trace back to ejection when the ejection is more compact, assuming the ejection forms on small scales comparable to the Schwarzschild radius (e.g. for AGN: \citealt{Janssen2021}), with a quasi-constant energy corresponding to the dashed blue line in Fig. \ref{fig:Wt}.
The MAXI J1820+070 ejection had a size $3.6\times 10^9 R_S < R < 1.0\times 10^{11} R_S$ at 90 days \citep{Bright_2020}, where $R_S \equiv \frac{2GM}{c^2}$.
An adiabatic expansion must eventually cease to be valid, as the internal energy is unbounded as $R \rightarrow 0$, hence we use Fig. \ref{fig:Wt} to justify using $W_{\text{eq}}\approx 10^{43}$ erg, as the energy derived from equipartition is roughly constant for the AMI-LA data at early times.
This value can be further motivated by considering an ejection produced at Eddington-limited power $L_{\text{Edd}}$ for the duration of the initial flare - similar to e.g. \cite{Sreehari2019}, although this may even be an underestimate of the internal energy $W_{\text{eq}}$ \citep{Cooper2025}.
For $W_{\text{eq}} \lesssim L_{\text{Edd}} t_{\text{flare}}$:
\begin{equation}
W_{\text{eq}} \lesssim 1.3\times10^{38} \frac{M_{\text{BH}}}{M_\odot} t_{\text{flare}} \implies W_{\text{eq}} \lesssim 2.6\times10^{43} \text{ erg}
\end{equation}
in good agreement with the value from equipartition in Fig. \ref{fig:Wt} for $R(90 \text{ days}) = 3.3\times10^3$ AU.

According to the flux as shown in Fig. \ref{fig:early gamma-rays}, $\gamma$-rays could have been detected with CTAO for hours if observed at ejection.
The Fermi-LAT measurement of no $\gamma$-rays at fluxes $\gtrsim 10^{-10}$ erg/cm$^{2}$/s \citep{Hoang2019,Abe2022} constrains the minimum possible radius to $\approx 5\times10^{6} R_S$ in our model.
For comparison, the earliest size measurement of the XTE J1908+094 ejection was $R\approx10^8 R_S$ \citep{Rushton2017} and the upper limit on the unresolved MAXI J1535-571 ejection was $\approx10^{10} R_S$ \citep{Russell_2019}, which is likely a large overestimate due to maximum energy considerations.
Hence, an early-time burst of $\gamma$-rays may be plausible to detect with a sensitive $\gamma$-ray observatory.
Notably, the $\gamma$-ray fluxes are comparable to the extended emission observed by \cite{LHAASO2024}, perhaps indicative of a continual interaction/build-up between ejecta and ISM around active BH-XRBs.
Further work is required in analysing the energy transfer between the two media, as in \cite{Savard2025}.

The detection of neutrinos is disfavoured by two considerations.
Firstly, neutrinos are only produced in appreciable quantities (when compared to the sensitivity of Trinity \citep{Wang_2021} or IceCube \cite{Aartsen_2019}) at sizes smaller than $10^4 R_S$, which is constrained by the Fermi-LAT upper limit according to our model.
A further constraint on the detection of neutrinos arises when considering the timescale of expansion $t_\text{exp}$: for $R\sim10^4 R_S$, $t_\text{exp}\approx 80$s, which severely limits the number of neutrino events observable.
Although not detectable from ejecta, neutrinos may be detectable from the extended emission regions as observed by \cite{LHAASO2024} if the above hypothesis of ejecta build-up is true.

\section{Conclusions}

In this work, we have examined whether discrete BH-XRB ejecta could be significant sources of cosmic rays, and be detectable as $\gamma$-ray or neutrino emitters, using MAXI J1820+070 as a case study. Our study allows us to make three theoretical predictions.
Firstly, ejecta from BH-XRBs may accelerate protons up to $\sim 10^{16}\mu^{-1/2}$ eV, making them plausible sites of extreme CR acceleration.
Secondly, BH-XRB ejecta could possibly contribute a flux ($\sim5\%$) to the knee of the CR spectrum, calculated here assuming that the energy share between non-thermal protons, non-thermal electrons and magnetic fields is not too dissimilar.
Deviations from equipartition may not rule out PeV CRs from BH-XRB ejecta, however the shape of the BH-XRB spectrum would change as a result; for instance, increasing $\mu$ to $100$ gives an upper limit of $\sim15\%$ at the knee.
Despite this being a conservative estimate, discrete ejecta from BH-XRBs appear to represent a only a fraction of observed PeV CRs and likely do not dominate at the knee.
Instead, the knee may be due to a combination of CRs from \textit{hard state} BH-XRB jets and/or other sources.
\cite{Cooper_2020} suggest that BH-XRB hard-state jets might be the dominant source of CRs at the knee, while accounting for only around a few percent of the total CR spectrum.
Moreover, \cite{Kaci2025} recently calculated BH-XRB population contribution to be $\gtrsim25\%$ at the knee, using the CR propagation code presented in \cite{Kaci2025a}.
Additionally, the PeV CRs could be dominated by either young, massive, windy star clusters \citep{Vieu2022,Aharonian2022} or skewed by a nearby PeV source.
Thirdly, $\gamma$-rays from the ejecta will be observed only if ejecta are formed on small scales and the Doppler boosting is favourable, and even then only as a short burst with a sensitive observatory like CTAO.
Detecting $>1$ TeV neutrinos from ejecta is highly unlikely, even if ejecta form on favourably small scales.
To continue this investigation, further simultaneous observations on differing angular scales permitting measurements of the size of ejecta are necessary.
This would allow ejecta and flare energies to be better correlated, and a Monte Carlo simulation of the overall BH-XRB contribution to the CR spectrum could be carried out as outlined.
In addition, the earlier these observations take place after ejection, the better we can understand the formation of these ejecta and determine whether TeV $\gamma$-rays are produced in measurable quantities or not.
Finally, considering the effect of propagation is necessary to refine the estimate of CR contribution from BH-XRBs.


\section*{Acknowledgements}

The authors thank Katie Savard, Laura Olivera-Nieto and Alicia López Oramas for helpful discussions.
AJC acknowledges support from the Oxford Hintze Centre for Astrophysical Surveys which is funded through generous support from the Hintze Family Charitable Foundation.
DK acknowledges funding from the French Programme d’investissements d’avenir through the Enigmass Labex, from the ‘Agence Nationale de la Recherche’, grant number ANR-19-CE310005-01 (PI: F. Calore), and from Tamkeen under the NYU Abu Dhabi Research Institute grant CASS.
JM acknowledges funding from a Royal Society University Research Fellowship (URF$\backslash$R1$\backslash$221062).

\section*{Data Availability}

Data will be made available upon request.
 



\bibliographystyle{mnras}
\bibliography{references} 




\appendix
\section{Normalisation of spectrum}
\label{app:integral}
There is a straightforward approximation for the integral.
\begin{equation}
 W_{\text{prot}} = \int_{E_{\text{min}}}^{\infty} N_0 E^{1-p} \exp \left({-\frac{E}{E_{\text{max}}}}\right) \,dE \approx \int_{E_{\text{min}}}^{E_{\text{max}}} N_0 E^{1-p}\,dE
\end{equation}
which is sufficient for our purposes.
Integrating gives
\begin{equation}
N_0 = (p-2) \frac{W_{\text{prot}}}{E_{\text{min}}^{2-p}-E_{\text{max}}^{2-p}}
\label{eq:Wq}
\end{equation}
We also form an equation for the total number of non-thermal protons $N_{\text{nth}}$:
\begin{equation}
N_{\text{nth}} = \int_{E_{\text{min}}}^{\infty} N_0 E^{-p} \exp \left({-\frac{E}{E_{\text{max}}}}\right) \,dE \approx \int_{E_{\text{min}}}^{E_{\text{max}}} N_0 E^{-p}\,dE
\end{equation}
which gives
\begin{equation}
N_0 = (p-1) \frac{N_{\text{nth}}}{E_{\text{min}}^{1-p}-E_{\text{max}}^{1-p}}
\label{eq:Nq}
\end{equation}
If we take Eq. \eqref{eq:Wq}/Eq. \eqref{eq:Nq}, and use $E_{\text{min}} \ll E_{\text{max}}$ we get:
\begin{equation}
E_{\text{min}} = \frac{p-2}{p-1} \times \frac{W_{\text{prot}}}{N_{\text{nth}}}
\end{equation}
We can use $N_{\text{nth}} = 1.0\times10^{44}$ \citep{Espinasse_2020} to find $E_{\text{min}}\approx 1.0\times10^{9}$ eV and $N_0 = 7.6\times 10^{54}$.
Numerical integration gives the same answers to 1 s.f.: $E_{\text{min}}\approx 1.1\times10^{9}$ eV and $N_0 = 5.3\times 10^{54}$.
This agrees with \cite{Kantzas2023a} as minimum non-thermal proton energy $\sim 1$ GeV.
Note that overall normalisation is rather insensitive to $E_\text{min}$: $N_0 \propto {E_\text{min}^{p-2}} = E_\text{min}^{0.2} \text{ for } p=2.2$.

\section{Expansion speed calculations for MAXI J1820+070}
\label{app:expansion}
\begin{table}
	\centering
	 \caption{Fitted proper expansion speeds for different sizes. Errors ($1 \sigma$) include both fitting and observation. The data in the second row corresponds to the solid purple and blue lines in Fig. \ref{fig:Wt}.}\label{tbl_exp}
	\begin{tabular}{cccc} 
		\hline
		$R(90 \text{ days})$ [AU]& $\beta_{\text{exp}}$ & $\beta_{\text{exp}}$ & $W_{\text{eq}}$ [$10^{42}$ erg] \\
		\hline
        $t=90$d & $t<80$d & $t > 80$d & $t \rightarrow 0$ \\
        \hline
 $6.2 \times 10^2$ & $0.0127\pm0.0003$ & $0.043\pm0.005$ & $(0.82\pm0.02)$ \\
 $3.3 \times 10^3$ & $0.064\pm0.003$ & $0.22\pm0.03$ & $(7.0\pm0.2)$\\
 $1.7 \times 10^4$ & $0.34\pm0.02$ & $1.2\pm0.1$ & $(59\pm2)$\\
		\hline
	\end{tabular}
\end{table}

The largest ejection size has an unphysical best-fit $\beta_{\text{exp}}$ at late times (see Table \ref{tbl_exp}), so larger observed sizes at $t=90$d should be viewed as more unlikely and so we estimate a robust cosmic ray maximum energy $E_\text{max} = 1.4\times10^{16}$ eV.
Regardless of parameter values, the best-fit expansion speed is $0.05 \lesssim \beta_{\text{exp}} \lesssim 0.25$, consistent with \cite{Rushton2017,Fender_2019}.
It takes years for the energy to change appreciably in the case of an adiabatic expansion from $\approx$ 90 days onwards, neglecting late-time energy transfers to the ISM.
After 2000 days, the internal energy decreases by a factor 6 for a typical $\beta_{\text{exp}} = 0.1c$.
Hence the state at $t = 90$ days is a pseudo-steady state, a convenient simplification when we calculate the $\gamma$-rays and neutrinos produced.
This breaks down for extreme expansion speeds ($\approx 0.3c$), as the internal energy decreases by roughly a factor 30 in the same time period, which would give a lower flux of $\gamma$-rays and neutrinos.
Additionally, fitting constant energy at early times ($t \lesssim 80$d) gives good agreement.
This may suggest that interactions with the environment become significant from this time onwards, perhaps due to an increase in ISM density.
This roughly constant energy motivates further early-time calculations in Section \ref{sec:NUG}.
Here we have neglected energy transfers between ejecta and the ISM, explored by \citet{Savard2025}.
X-ray and radio measurements suggest that the ejection from MAXI J1820+070 decelerates at later times \citep{Espinasse_2020}, which could correspond to it reaching the edge of a density cavity the BH-XRB is located in \citep{Rushton2017,Espinasse_2020,Carotenuto_2022}.

\section{Steady-state production of gamma-rays and neutrinos}
\label{app:steady state}
\begin{figure}
\centering
\includegraphics[width=\columnwidth]{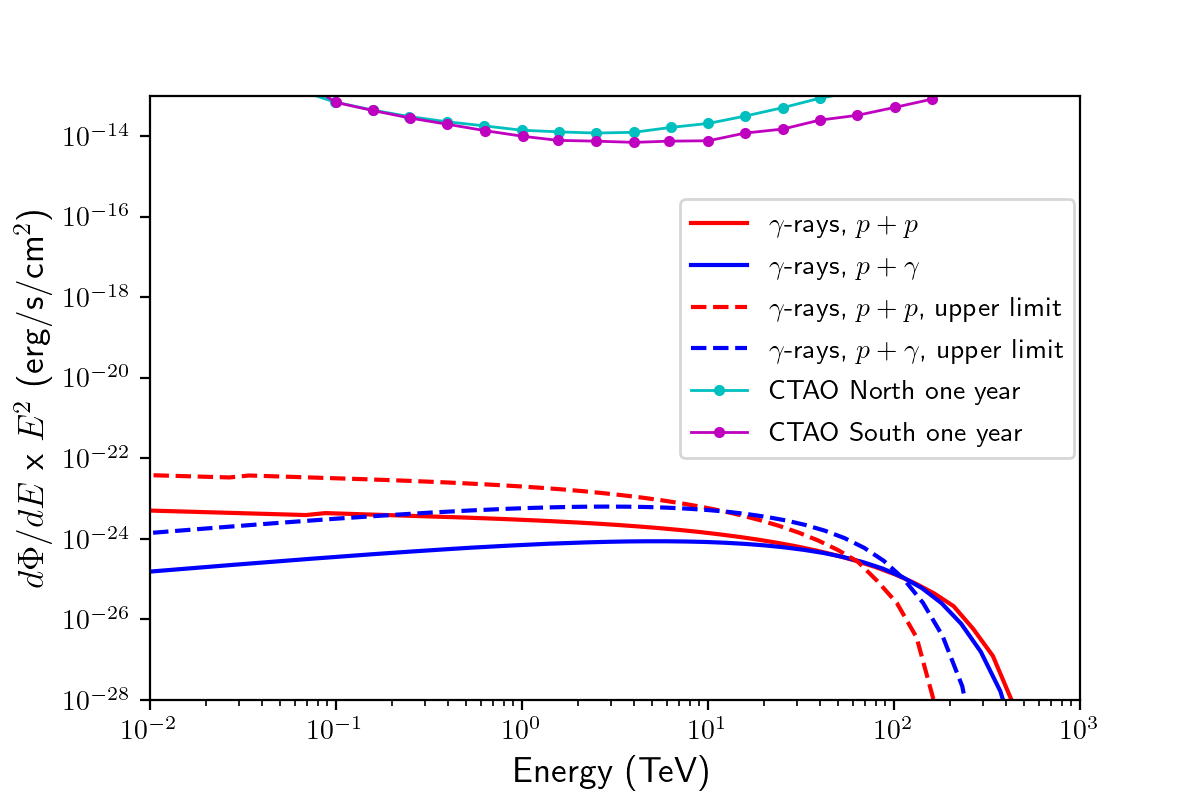}
\caption{Rate of $\gamma$-rays produced from the MAXI J1820+070 ejection. TeV $\gamma$-rays are undetectable. Predicted CTAO sensitivity \citep{Observatory2021} has been plotted for comparison.} 
\label{fig:gamma-rays}
\end{figure}

\begin{figure}
\centering
\includegraphics[width=\columnwidth]{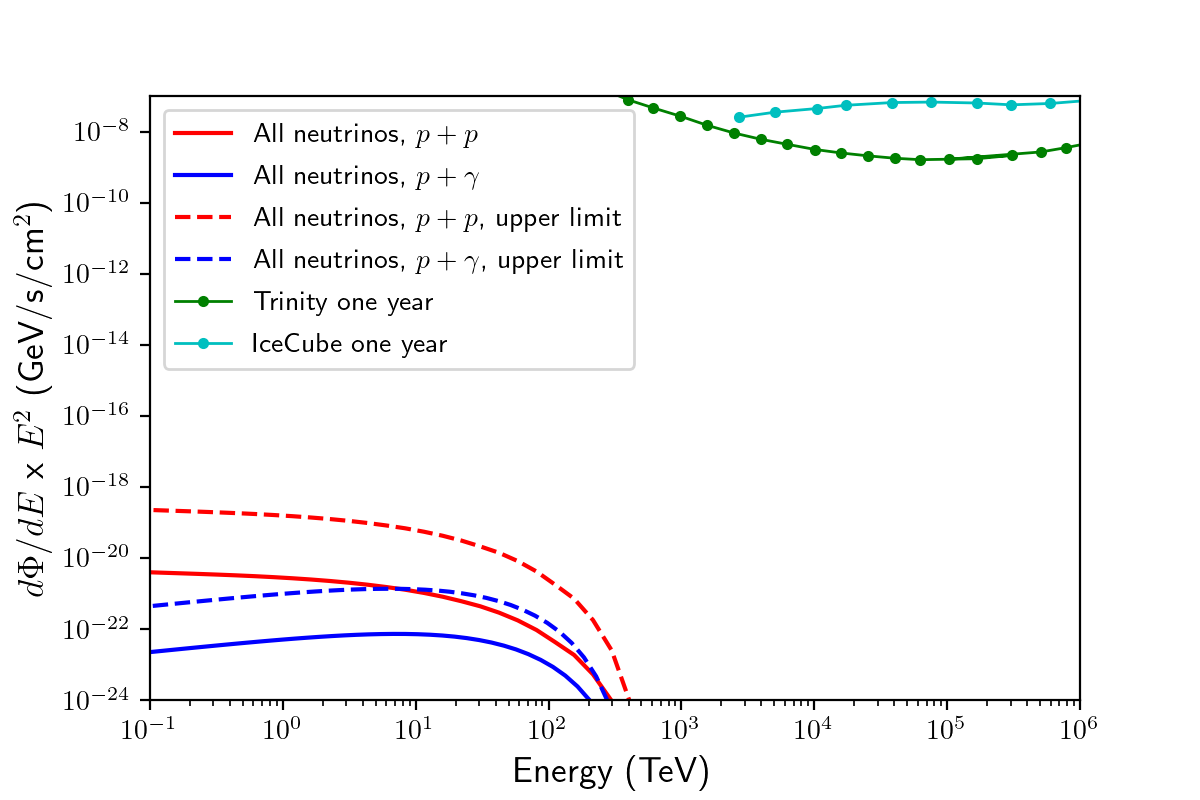}
\caption{The calculated rate of neutrino production from the MAXI J1820+070 ejection is orders of magnitude below sensitivity of detectors.} 
\label{fig:neutrinos}
\end{figure}

Rate of production of $\gamma$-rays (Fig. \ref{fig:gamma-rays}) and neutrinos (Fig. \ref{fig:neutrinos}) from the MAXI J1820+070 ejection onwards from $t=90$d in the pseudo-steady state.
Sensitivity scales as $\sqrt{t_\text{obs}}$ for long exposures and event counts $N>>1$.

\bsp	
\label{lastpage}
\end{document}